\documentclass[aps,preprint]{revtex4}
\usepackage{amsmath}
\usepackage{amssymb}
\usepackage{mathrsfs}
\usepackage{xcolor}
\usepackage{graphicx}
\usepackage{subfigure}
\usepackage{enumerate}
\usepackage{float}

\begin{document}
	%\preprint{CTP-SCU/2021010}	
	\title{Gravitational Lensing by Black Holes in Einstein-nonlinear Electrodynamic Theories with Multiple Photon Spheres}

	\author{Siyuan Hui$^{a,b}$}
	\email{huisiyuan@stu.scu.edu.cn}
	\author{Benrong Mu$^{a,b}$}
	\email{benrongmu@cdutcm.edu.cn}
	\thanks{Corresponding author}
	\author{Peng Wang$^{b}$}
	\email{pengw@scu.edu.cn}
	
	\affiliation{$^{a}$Center for Joint Quantum Studies, College of Medical Technology, Chengdu University of Traditional Chinese Medicine, Chengdu, 611137, PR China\\
		$^{b}$Center for Theoretical Physics, College of Physics, Sichuan University, Chengdu, 610064, PR China}

	\begin{abstract}
	
	In this paper, we study the gravitational lensing effects of non-linear electrodynamic black holes. Non-linear electrodynamic black holes serve as typical models for multi-event horizon black holes. Depending on the choice of metric parameters, these black holes can possess more than five event horizons. Consequently, within certain parameter ranges, black holes can have more than three photon spheres of varying sizes outside the event horizon. Specifically, we focus on the strong gravitational lensing effects near the triple photon spheres, particularly the formation of higher-order images of point sources and celestial spheres. The presence of one, two, or three or more photon spheres significantly increases the number of higher-order images of a point source. When a black hole is illuminated by a celestial sphere, the three photon spheres generate three critical curves in the black hole image, with the smallest critical curve coinciding with the shadow's edge. Additionally, since non-linear electrodynamic black holes are models of multi-event horizon black holes, we can infer the gravitational lensing effects and the changes in celestial images for black holes with more than three photon spheres by analyzing the distinctions and patterns between the gravitational lensing effects of one, two, and three photon spheres.

	\end{abstract}
	\maketitle
	\tableofcontents
	
	\section{Introduction}
	
	One of the most profound predictions of general relativity is the deflection of light in curved spacetime, known as gravitational lensing. Gravitational lensing has become a crucial tool for addressing fundamental questions in astrophysics and cosmology. The study of gravitational lensing in the strong-field regime, particularly near compact objects such as black holes and neutron stars, began in earnest in the 1970s \cite{Luminet:1979nyg}. Since then, extensive research has explored gravitational lensing caused by various structures \cite{Mellier:1998pk, Bartelmann:1999yn, Heymans:2013fya}, including dark matter \cite{Kaiser:1992ps, Clowe:2006eq, Atamurotov:2021hoq}, dark energy \cite{Biesiada:2006zf, Cao:2015qja, DES:2020ahh, DES:2021wwk, Zhang:2021ygh}, quasars \cite{SDSS:2000jpb, Peng:2006ew, Oguri:2010ns, Yue:2021nwt}, gravitational waves \cite{Seljak:2003pn, Diego:2021fyd, Finke:2021znb}, and other compact objects \cite{Schmidt:2008hc, Guzik:2009cm, Liao:2015uzb, Goulart:2017iko, Nascimento:2020ime, Junior:2021svb, Islam:2021ful, Tsukamoto:2022vkt}. Recently, the Event Horizon Telescope collaboration achieved the angular resolution necessary to capture the image of the supermassive black hole at the center of galaxy M87, marking a new era in the study of gravitational lensing within the strong gravity regime \cite{EventHorizonTelescope:2019dse, EventHorizonTelescope:2019uob, EventHorizonTelescope:2019jan, EventHorizonTelescope:2019ths, EventHorizonTelescope:2019pgp, EventHorizonTelescope:2019ggy, EventHorizonTelescope:2021bee, EventHorizonTelescope:2021srq}. In particular, the black hole shadow observed in these images is closely related to the properties of photon spheres and strong gravitational lensing effects near the event horizon \cite{Falcke:1999pj, Virbhadra:1999nm, Claudel:2000yi, Eiroa:2002mk, Virbhadra:2008ws, Yumoto:2012kz, Wei:2013kza, Zakharov:2014lqa, Atamurotov:2015xfa, Cunha:2016wzk, Dastan:2016bfy, Amir:2017slq, Wang:2017hjl, Ovgun:2018tua, Perlick:2018iye, Kumar:2019pjp, Zhu:2019ura, Ma:2019ybz, Mishra:2019trb, Zeng:2020dco, Zeng:2020vsj, Qin:2020xzu, Saurabh:2020zqg, Roy:2020dyy, Li:2020drn, Kumar:2020hgm, Zhang:2020xub, Olmo:2021piq, Guerrero:2022qkh, Virbhadra:2022iiy}.
	
	Interestingly, certain special black holes have been identified as having two photon spheres outside the event horizon within specific parameter regions. Examples include scalarized hairy black holes \cite{Gan:2021pwu, Gan:2021xdl}, dyonic black holes with a quasitopological electromagnetic term \cite{Liu:2019rib}, and black holes in massive gravity \cite{deRham:2010kj, Dong:2020odp}. The presence of two photon spheres can substantially influence the optical appearance of black holes illuminated by surrounding accretion disks, potentially producing bright rings of varying radii in black hole images and significantly enhancing the flux of the observed images. Furthermore, the effective potential for scalar perturbations in black holes with two photon spheres has been shown to exhibit a double-peak structure, which can give rise to long-lived quasinormal modes and echo signals \cite{Guo:2021enm, Guo:2022umh, Huang:2021qwe, Hui:2023ibl}.
	
	In this paper, we examine general Einstein non-linear electrodynamic theories. These theories were first introduced by Born and Infeld, known as the Born-Infeld theory, to address the problem of infinite self-energy associated with point charges \cite{Chakravarti:2021clm, Dong:2020odp}. In the context of gravitational theory, incorporating non-linear electromagnetic fields may offer a solution to avoid black hole singularities \cite{Huang:2021qwe, Guo:2022umh, Li:2019kwa, Gao:2021kvr, Born:1934ji}. Furthermore, recent studies suggest that non-linear electrodynamic theories can lead to multiple horizons and critical points \cite{Born:1934gh, Bambi:2013caa}. Due to the high degrees of freedom inherent in these theories, it is possible to select an appropriate coupling constant, resulting in a black hole’s effective potential exhibiting multiple peaks. In such scenarios, these peaks can emerge far from the event horizon, representing more than just a correction to the near-horizon regime. Recently, the time-domain profile of scalar perturbations in black holes has been computed, revealing that three-peak profiles differ significantly from two-peak profiles \cite{Sang:2022hng}. When the number of peaks increases to three, the wave packet undergoes splitting. This remarkable phenomenon allows us to infer the number of peaks in a black hole potential formed after a binary black hole merger, based on gravitational wave signals.
	
	Gravitational lensing and echo effects play crucial roles in the observation of black holes, and there have been studies discussing the impact of dual photon spheres on black hole gravitational lensing \cite{Guerrero:2022qkh, Guo:2022muy}. However, the influence of three or more photon spheres on black hole gravitational lensing has yet to be explored. The effective potentials with three or more peaks have shown significant differences from those with two peaks in relation to the black hole echo effect. In this paper, we aim to investigate whether the impact of black holes with three or more photon spheres on gravitational lensing is as pronounced as the multi-peak echo effect, and whether there exists any underlying patterns. The remainder of this paper is organized as follows. In Sec.\ref{222}, we briefly review the metric of black holes in Einstein-nonlinear Electrodynamic Theories and give out the method for ray tracing. In Sec.\ref{333}, we give out the results for each situation and summarize the transformation rules for black holes with multiple photon spheres. In Sec.\ref{444}, the conclusion is given. We set $G=c=1$ through out the paper.

	\section{Solutions for Black Holes in Einstein-nonlinear Electrodynamic Theories}\label{222}
	The general action of the Einstein gravity minimally coupled with the non-linear electromagnetic field can be written as
	\begin{equation}\label{1}
		S=\int d^4x \sqrt{-g}(R+L_{EM}),
	\end{equation}
	where
	\begin{equation}\label{2}
		\begin{aligned}
			&L_{EM} = \sum_{i=1}^{\infty} \alpha_i {(F^2)}^{-i},\\
			&F^2 = F_{\mu \nu} F^{\mu \nu},\\
			&F_{\mu \nu} = \nabla_{\mu} A_{\nu} - \nabla_{\nu} A_{\mu}.
		\end{aligned}
	\end{equation}
	Here, $R$ is the Ricci scalar and $L_{EM}$ is the extended Maxwell Lagrangian, $A_{\mu}$ is the Maxwell field, $\alpha_i$ are dimensional constants and $i$ is a positive integer \cite{Gao:2021kvr}. When $\alpha_1=1$ and $\alpha_i=0$ for any $i > 1$, the theory reduces to the Einstein-Maxwell theory. The variation of Eq. (\ref{1}) with respect to $g_{\mu \nu}$ and $A_\mu$ gives the equation of motion
	\begin{equation}\label{3}
		\begin{aligned}
			&G_{\mu \nu}= -2 \frac{\partial L_{EM}}{\partial F} F_{\mu}^{\lambda}F_{\nu \lambda} + \frac{1}{2} g_{\mu \nu} L_{EM},\\
			&\nabla_{\mu} \left(\frac{\partial L_{EM}}{\partial F} F^{\mu \nu} \right) =0.
		\end{aligned}
	\end{equation}
	The general asymptotically flat black hole solution can be written as
	\begin{equation}\label{4}
		ds^2 = -f(r)dt^2 +\frac{1}{f(r)} dr^2 + r^2 (d\theta^2 + \sin ^2 \theta d\phi^2),
	\end{equation}
	where
	\begin{equation}\label{5}
		\begin{aligned}
			f(r) =1 + \sum_{i=1}^{\infty} c_i r^{-i}, \quad
			A_t(r) = \sum_{i=1}^{\infty} b_i r^{-i}.
		\end{aligned}
	\end{equation}
	Here, $c_i$ and $b_i$ are constant parameter related to $\alpha_i$, the charge $b_1 = Q$ and the ADM mass $c_1=-2M$, and the relationship between the black hole parameters and the coupling constant can refer to \cite{Gao:2021kvr}.
	
	Light rays propagating in the black hole spacetime are described by the geodesic equations
	\begin{equation}\label{6}
		\frac{d^2 x^{\mu}}{d \lambda^2} + \Gamma^{\mu}_{\rho \sigma} \frac{d x^{\rho}}{d \lambda} \frac{d x^{\sigma}}{d \lambda} =0,
	\end{equation}
	where $\lambda$ is the affine parameter, and $\Gamma^{\mu}_{\rho \sigma}$ is the Christoffel symbol. Using $ds^2 = 0$, one can rewrite the radial component of the geodesic equations as
	\begin{equation}\label{7}
		\frac{1}{b^2} = \left( \frac{d r}{d \lambda} \right)^2 + \frac{f(r) }{r^2}
	\end{equation}
	where we rescale the affine parameter $\lambda$ as $\lambda/\lvert L \lvert$, $b=\lvert L \lvert/E$ is the impact parameter, and L and E are the conserved angular momentum and energy of photons, respectively. From Eq. (\ref{6}), the effective potential governing light rays is defined as
	\begin{equation}\label{8}
		V_{eff} (r) =\frac{f(r) }{r^2}
	\end{equation}
	Unstable circular null geodesics at radius $r_{ph}$, which constitute a photon sphere of radius $r_{ph}$, are determined by
	\begin{equation}
		\begin{aligned}
			V_{eff} (r_{ph}) =\frac{1}{b_{ph}^2}, \quad
			V'_{eff} (r_{ph}) = 0, \quad
			V''_{eff} (r_{ph}) < 0.
		\end{aligned}
	\end{equation}
	where $b_{ph}$ is the corresponding impact parameter. The effective potential $V_{eff} (r)$ of the aforementioned black hole solutions is plotted in Fig. 1. Remarkably, it shows that, when the black hole parameters take suitable values, $V_{eff} (r)$ can have multipal local maxima, leading to a multipal-peak structure. Consequently, black holes with the mutiple-peak effective potential possess mutipal photon spheres outside the event horizon.
%	\begin{figure}[htbp]
%		\centering
		%\subfigure[observer $O$]{
		%	\includegraphics[scale=1]{}}
%		\caption{Black hole solutions and associated effective potentials as a function of $r$ in Einstein-nonlinear Electrodynamic Theories.}
%	\end{figure}

	\section{Numerical Results}\label{333}
	To illustrate gravitational lensing by black holes in Einstein-nonlinear electrodynamic theories with single-peak and multipal-peak effective potentials, we place a luminous celestial sphere at $r_{CS} = 12 r_h$. Here, we choose different parameters to keep the radius of event horizon to be $r_h = 1$. The celestial sphere is concentric with the black holes and encloses observers. Two static observers, henceforth denoted as $O$ and $P$, are located at $x^{\mu}_O = (0,6 M,\pi/2,0)$ and $x^{\mu}_P = (0,6 M,\pi/3,0)$, respectively, and their viewing angles capture $2\pi/3$ of the celestial sphere. From the position of the observer, we scan their viewing angles by numerically integrating $4000 \times 4000$ null geodesics until reaching the celestial sphere or hitting the event horizon. The light source on the celestial sphere is divided into four quadrants, each of which is painted with a different color. Moreover, we also lay a set of black lines of constant longitude and latitude, with adjacent lines separated by $\pi/180$ \cite{Guo:2022muy}.

	\subsection{Single-peak potential}
	
	\begin{figure}[htbp]
		\centering
		\includegraphics[scale=0.68]{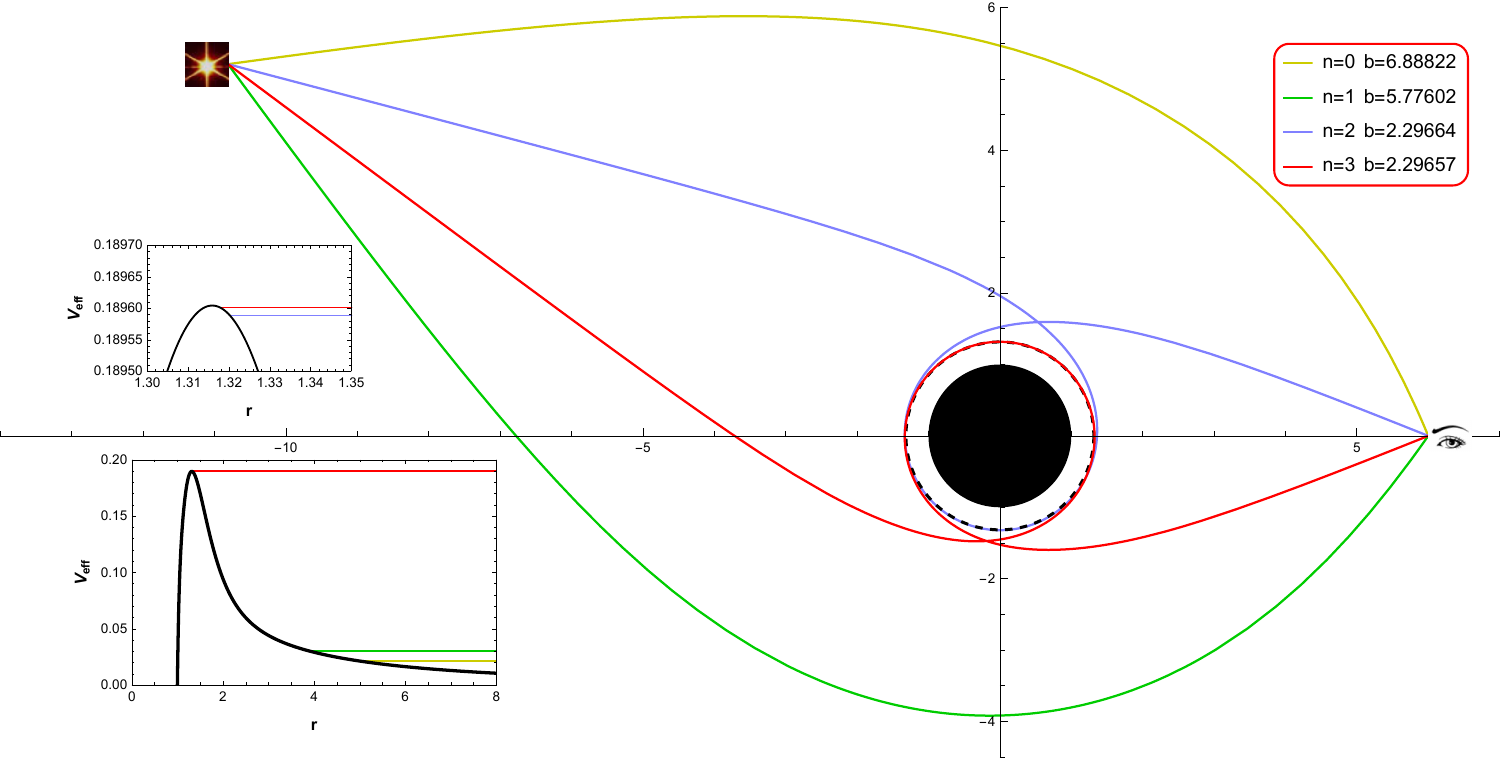}
		\caption{Light rays producing n-order images of a point like light source on the equatorial plane of the black hole.}
	\end{figure}
	
	\begin{figure}[htbp]
		\centering
		\subfigure[\quad Observer $O$]{
			\includegraphics[scale=0.07]{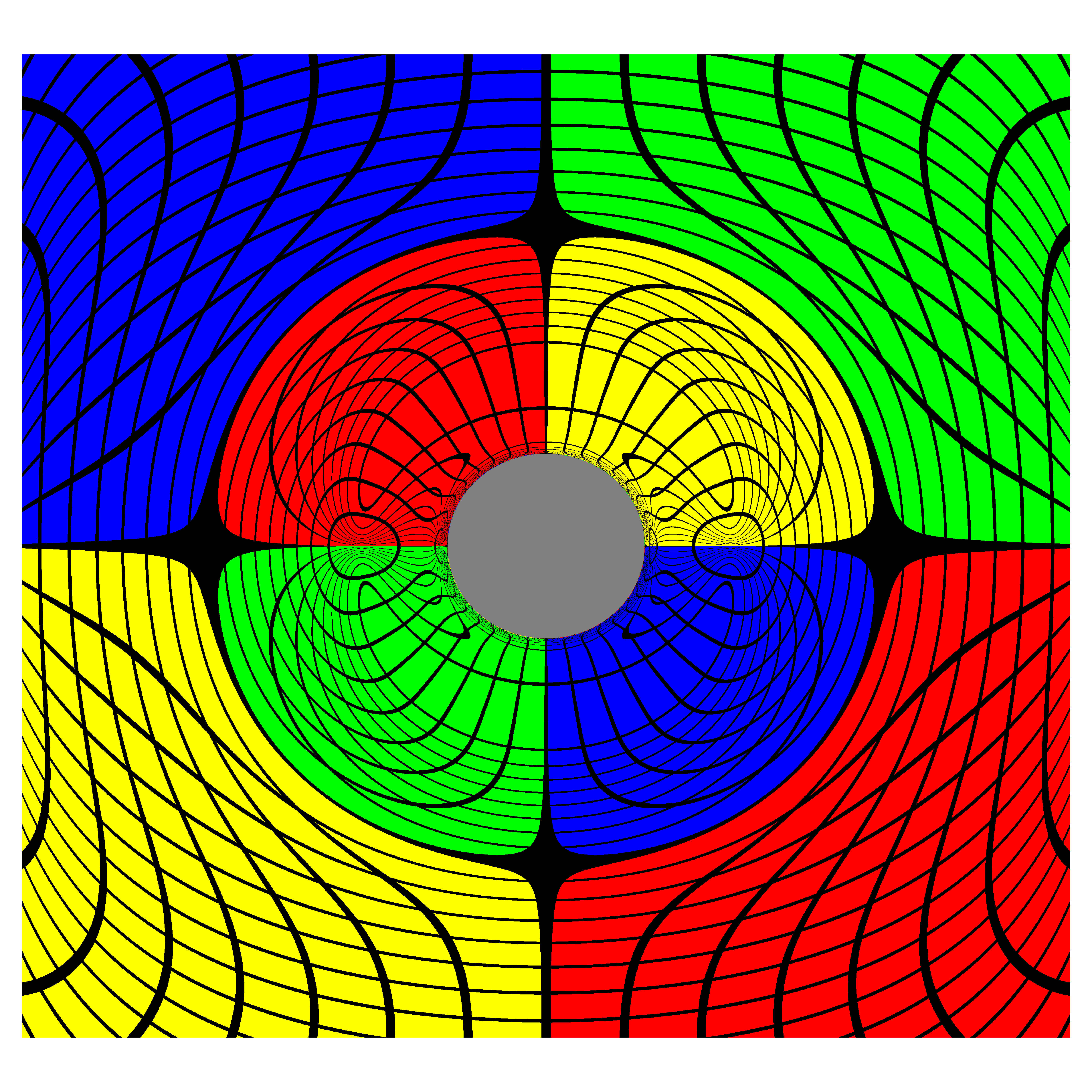}}
		\quad
		\subfigure[\quad Observer $P$]{
			\includegraphics[scale=0.07]{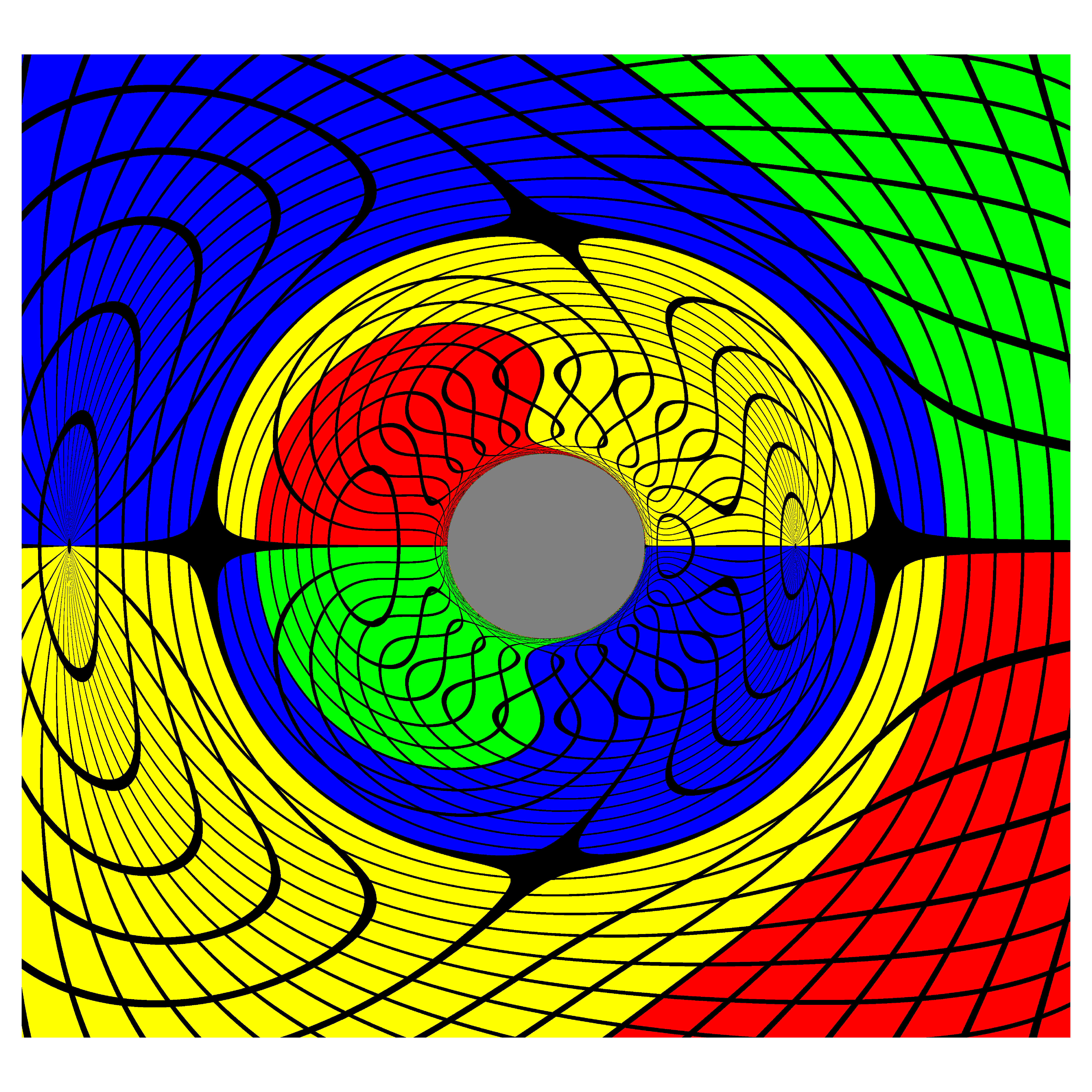}}
		\caption{Images of the black hole with single photon sphere, which is illuminated by the celestial sphere, viewed by the observers $O$ (left) and $P$ (right).}
	\end{figure}

	Here, we first consider the case where the black hole has a single-peaked effective potential, as illustrated in Figure 1. A single-peaked effective potential corresponds to the existence of a single photon sphere outside the event horizon, which is a closed unstable circular orbit for photons. The presence and number of photon spheres play a critical role in determining the structure of strong gravitational lensing and the formation of higher-order images. In Figure 1, we plot the light rays that connect point sources on the celestial equator to an observer $O$ under different impact parameters. It is known that, for any gravitational field with a single-peaked effective potential, the impact parameter determines how light rays are deflected in such a gravitational field. The impact parameter reaches its critical minimum value at the peak of the effective potential. 
	From Figure 1, we can observe that as the impact parameter decreases and approaches the critical value, the light rays are increasingly influenced by the proximity to the photon sphere, resulting in the formation of higher-order images with greater frequency. Ultimately, when the impact parameter falls below the critical value, light rays emitted by the point source are captured by the gravitational field of the black hole and eventually fall into the black hole. This is reasonable and explains why, in the context of black hole gravitational lensing, the observer can only receive information from light rays that originate outside the radius of the photon sphere of the black hole.
	In Figure 2, we illustrate the image of the black hole as seen by observers $O$ and $P$ when the celestial sphere illuminates the black hole. The gray region in the image represents the black hole's shadow, with its boundary being the critical curve, also referred to as the apparent boundary by Bardeen. Similar to the point source case, higher-order images of the celestial sphere appear outside the critical curve, becoming increasingly dense as they approach the critical curve. These densely packed higher-order celestial images constitute the photon ring \cite{Gralla:2019xty}.

	\subsection{Double-peak potential}
	
	\begin{figure}[htbp]
		\centering
		\includegraphics[scale=0.68]{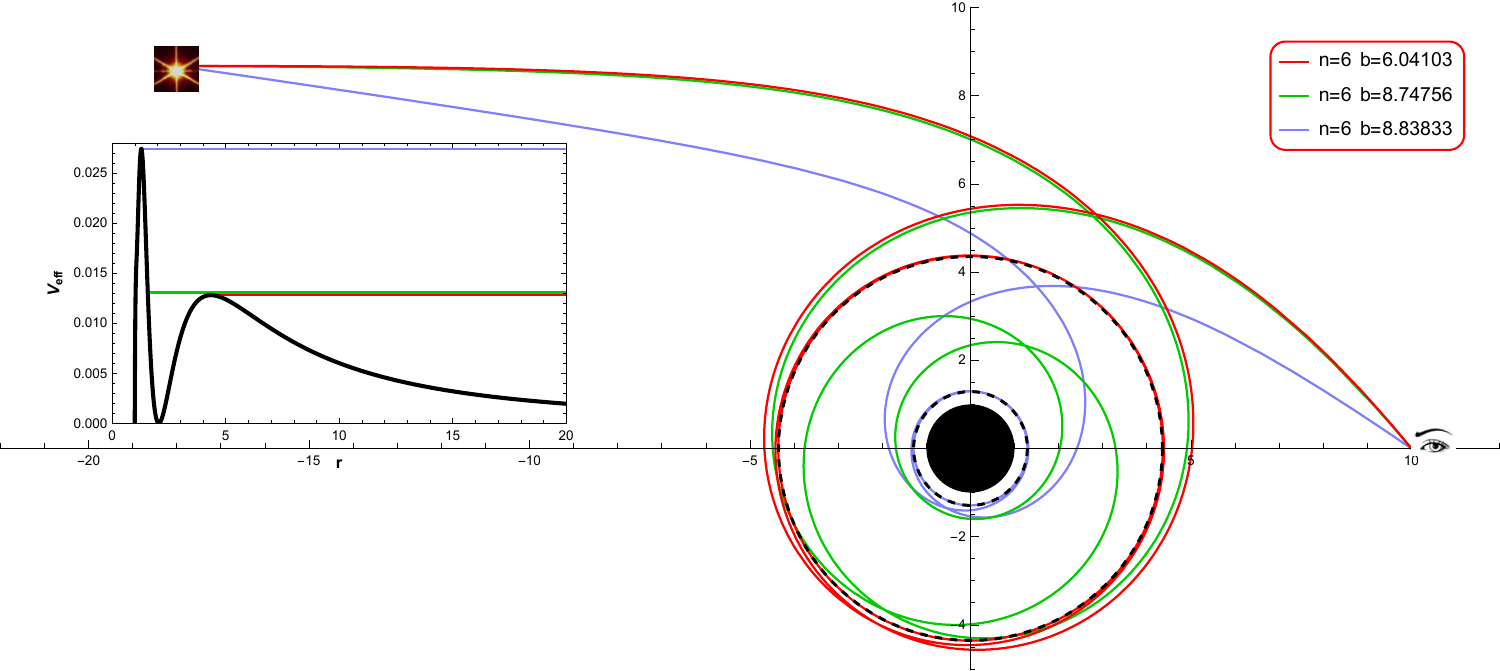}
		\caption{Light rays producing n-order images of a point like light source on the equatorial plane of the black hole.}
	\end{figure}
	
	\begin{figure}[htbp]
		\centering
		\subfigure[\ Observer $O$]{
			\includegraphics[scale=0.05]{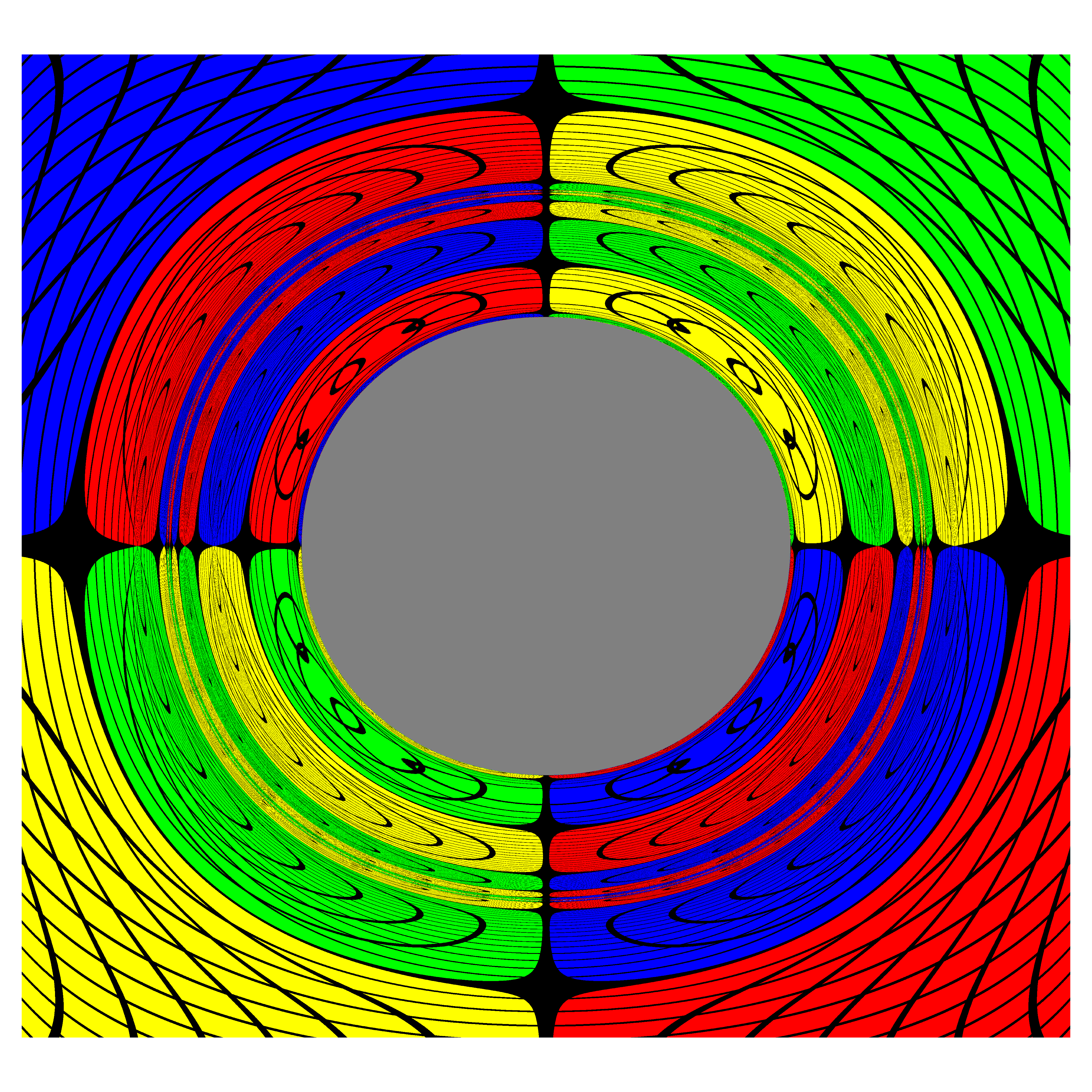}}
		\quad
		\subfigure[\ Observer $P$]{
			\includegraphics[scale=0.05]{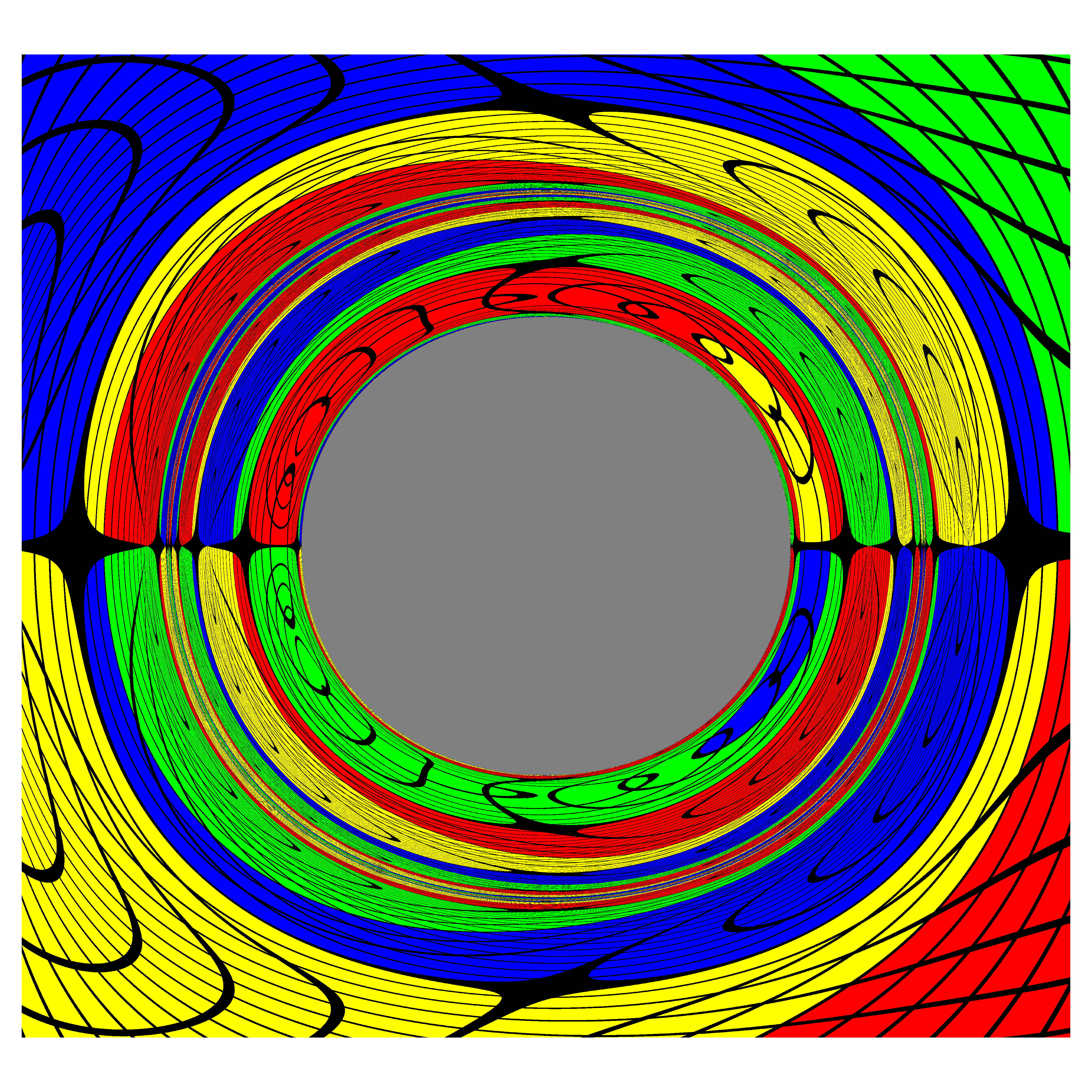}}
		\caption{Images of the black hole with double photon spheres, which is illuminated by the celestial sphere, viewed by the observers $O$ (left) and $P$ (right). Here, the inner peak is larger than the outer one.}
	\end{figure}
	
	\begin{figure}[htbp]
		\centering
		\subfigure[\ Observer $O$]{
			\includegraphics[scale=0.05]{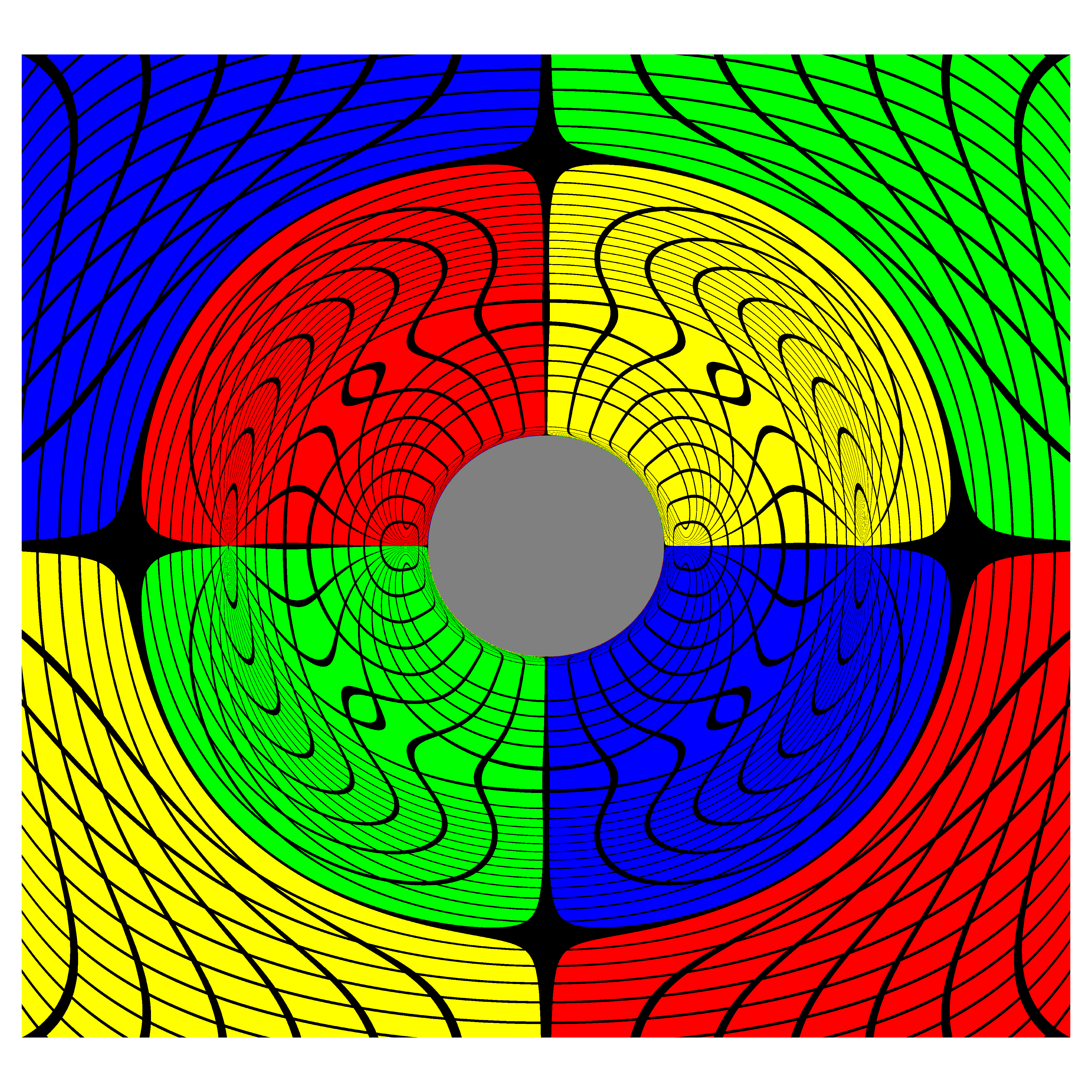}}
		\quad
		\subfigure[\ Observer $P$]{
			\includegraphics[scale=0.067]{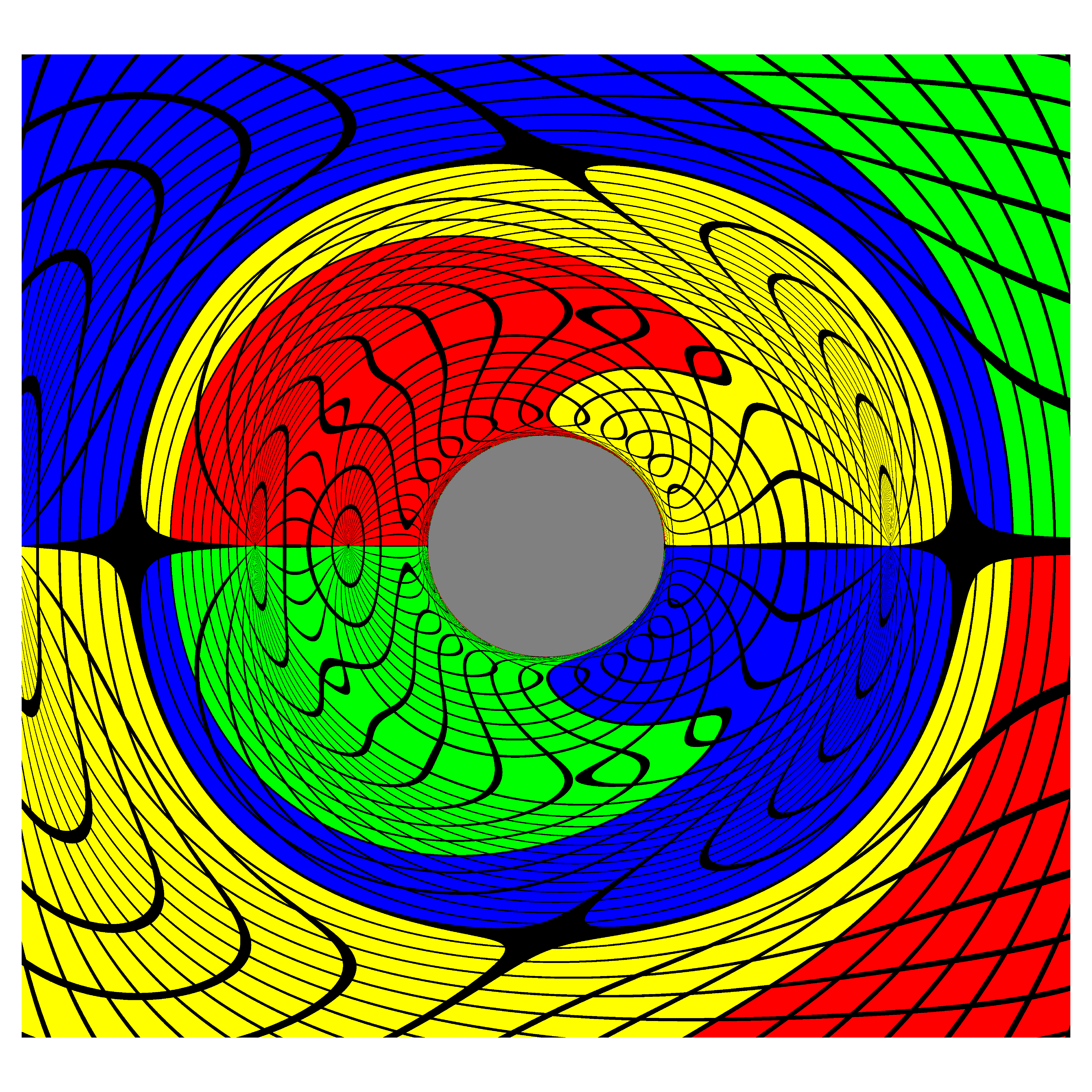}}
		\caption{Images of the black hole with double photon spheres, which is illuminated by the celestial sphere, viewed by the observers $O$ (left) and $P$ (right). Here, the inner peak is smaller than the outer one.}
	\end{figure}
	
	In Figures 3, 4 and 5, we discuss the scenario where the black hole has a double-peaked effective potential. The two peaks of the effective potential correspond to two photon spheres located outside the event horizon of the black hole. The relative heights of these peaks determine the distribution and visibility of higher-order images, as well as the appearance of multiple bright photon rings in the observed images.
	As shown in Figure 3 and 4, when the peak value of the inner effective potential is higher than that of the outer one, we need to pay attention to the paths of light rays under different impact parameters during ray tracing. Notably, in this scenario, for a given \( n \), the observer can see multiple higher-order images. In contrast, for a point source in a black hole with a single-peaked effective potential, only one higher-order image can be produced for a given \( n \). It is known that the value of \( n \) corresponds to the degree of deflection of the light ray, which is influenced by the effective potential. Therefore, we can observe that, apart from the case corresponding to the outer peak, which is consistent with the single-peaked situation, there exist certain impact parameters where the light ray is first deflected by the weaker influence of the outer peak and enters the outer photon sphere. It then undergoes a stronger influence from the inner peak, experiencing multiple significant deflections up to the \( n \)-th order, before being received by the observer.
	If the light source is the celestial sphere, then strong gravitational lensing near the double photon spheres can leave unique imprints on the black hole's image. Specifically, due to the combined influence of the double photon spheres, higher-order images of the celestial sphere will accumulate near the critical curve. In the strong deflection limit, these higher-order images asymptotically approach the critical curve, consistent with the behavior observed in the Hairy black hole with double photon spheres \cite{Guo:2022muy}.
	
	As shown in Figure 5, when the peak value of the inner effective potential is lower than that of the outer peak, we can observe that the outer effective potential is so strong that once the light ray is forcibly deflected into the photon sphere, the strong outer effective potential prevents the light ray from deflecting out again and being received by the observer. This explains why, in the celestial images, we see a similarity to the image produced by a single photon sphere associated with a single-peaked effective potential. Therefore, in this case, the influence of the inner peak is completely masked by the outer peak, resulting in a gravitational lensing image that externally appears to exhibit a single-peaked configuration.

	\subsection{Triple-peak potential}
	
	\begin{figure}[htbp]
		\centering
		\includegraphics[scale=0.68]{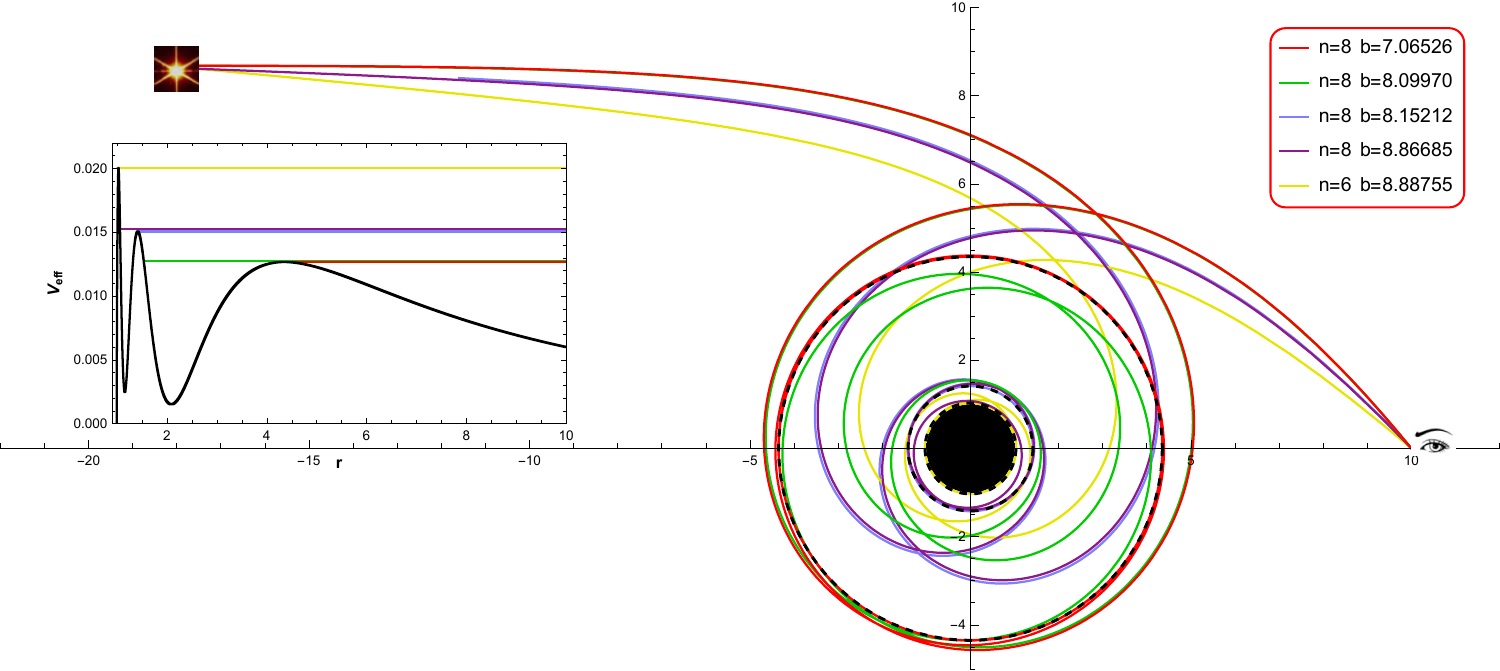}
		\caption{Light rays producing n-order images of a point like light source on the equatorial plane of the black hole.}
	\end{figure}
	
	\begin{figure}[htbp]
		\centering
		\subfigure[\ Observer $O$]{
			\includegraphics[scale=0.067]{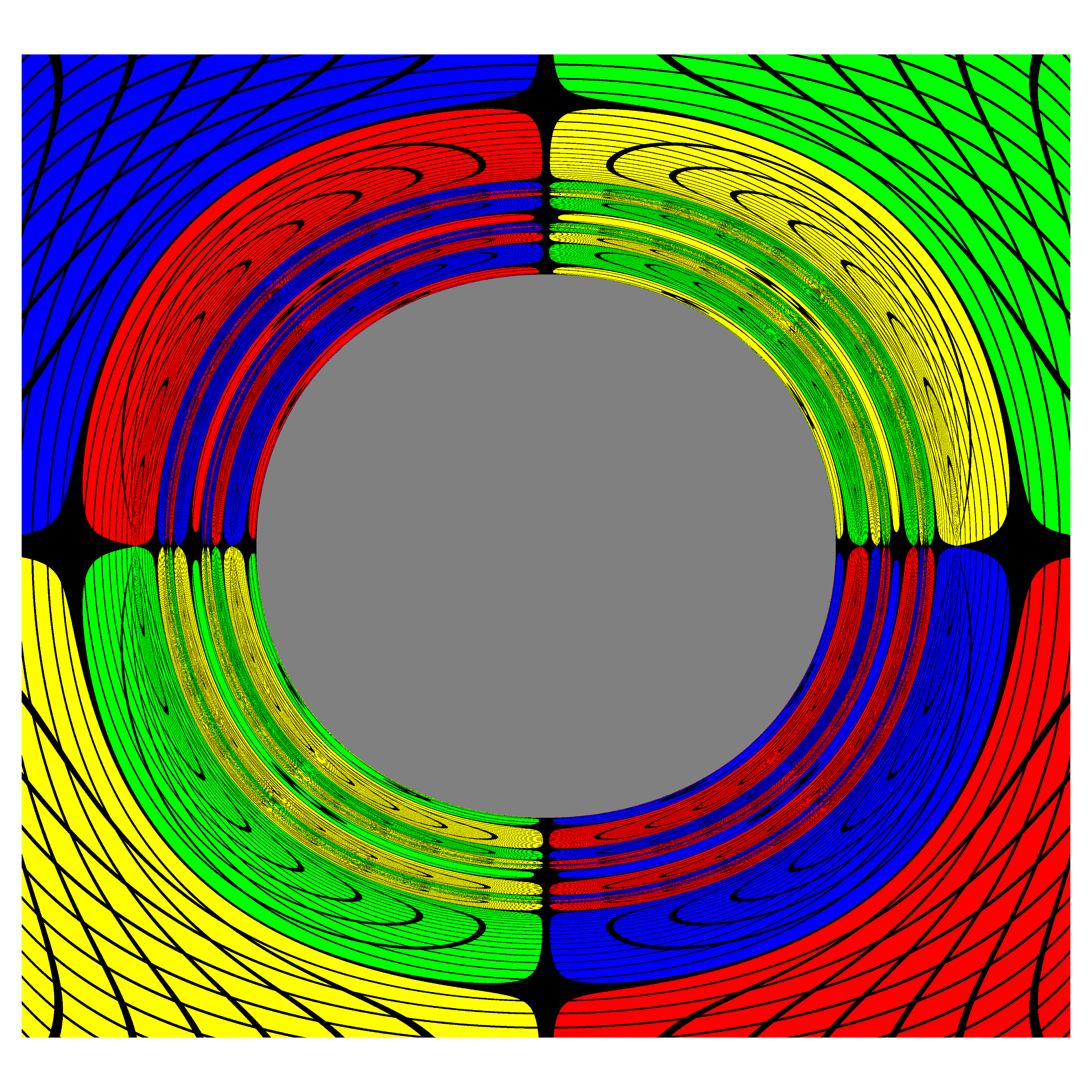}}
		\quad
		\subfigure[\ Observer $P$]{
			\includegraphics[scale=0.05]{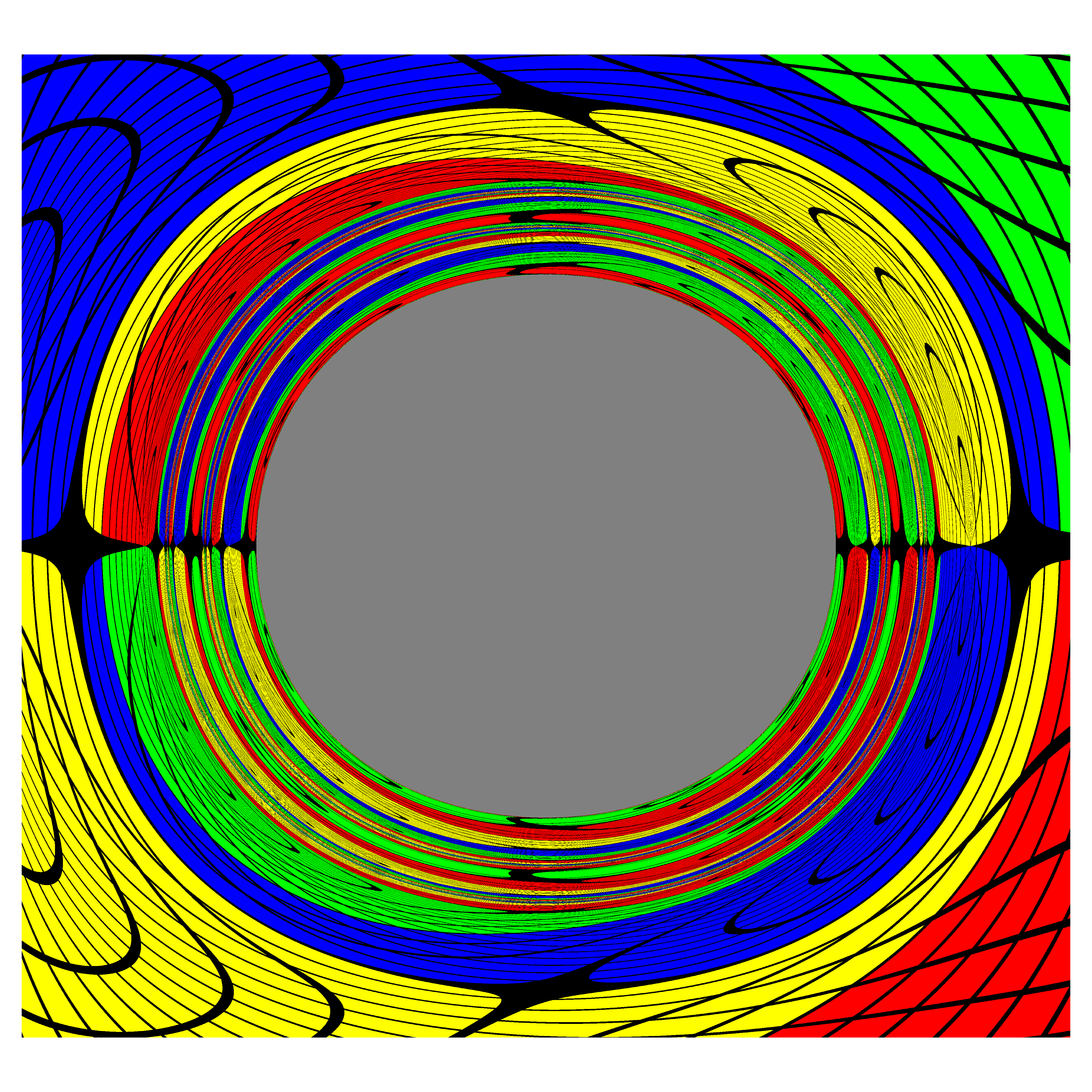}}
		\caption{Images of the black hole with triple photon spheres, which is illuminated by the celestial sphere, viewed by the observers $O$ (left) and $P$ (right). Here, the inner peak is the largest one, the outer peak is the smallest one.}
	\end{figure}
	
	As shown in Figure 6 and 7, generally, for a black hole with three peaks in its effective potential, these peaks decrease gradually from the inside out, with the innermost peak being the highest and the outermost peak being the lowest. In this scenario, the behavior of light rays retains some similarity to the double-peaked case, with photons potentially orbiting around multiple photon spheres. However, the presence of a third peak introduces additional complexity, resulting in a richer structure of higher-order images and more intricate accumulation patterns near each critical curve. These rays are affected by multiple peaks, causing them to rotate around these peaks multiple times, thereby completely altering their trajectories. In this case, we observe that the number of higher-order gravitational lensing images produced by a triple-peaked black hole far exceeds that of a double-peaked black hole.
	In general regions, the behavior of photons is similar to that seen in double-peaked and single-peaked configurations. However, in special regions, the number of higher-order images can fluctuate significantly with changes in the photon's impact parameter. As we gradually reduce the magnitude of one or two of the peaks, the photon paths start to resemble those seen in the double-peaked or single-peaked scenarios. This observation aligns with our expectations for the gravitational lensing effects of a triple-peaked black hole: the more peaks there are, the more complex the accumulation of higher-order images around the photon spheres will be. Moreover, the accumulated higher-order images tend to exhibit a certain pattern, with the images layered in sequence, extending outward from the photon sphere.

	\subsection{Special peak cases}
	
	\begin{figure}[htbp]
		\centering
		\subfigure[\ Observer $O$]{
			\includegraphics[scale=0.05]{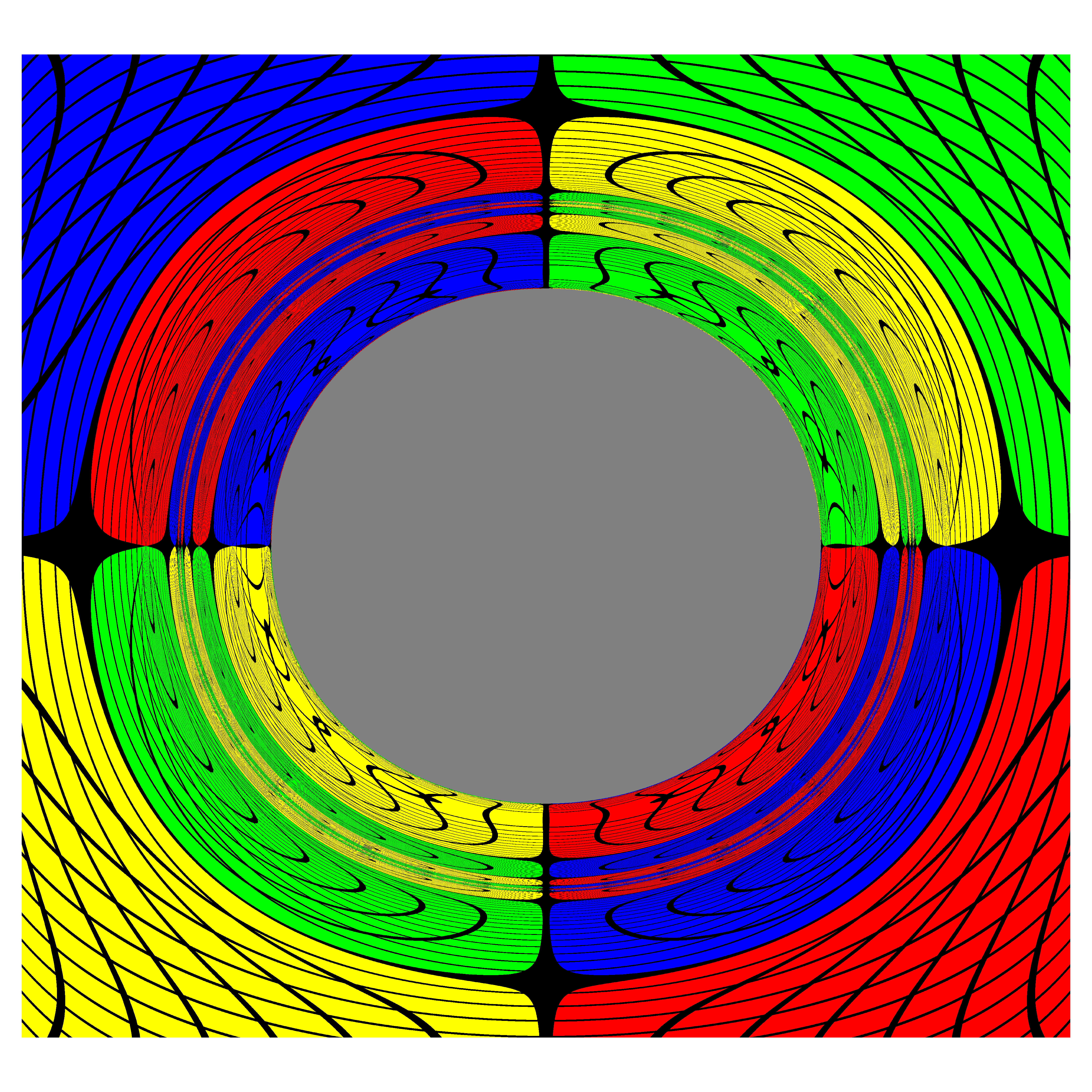}}
		\quad
		\subfigure[\ Observer $P$]{
			\includegraphics[scale=0.05]{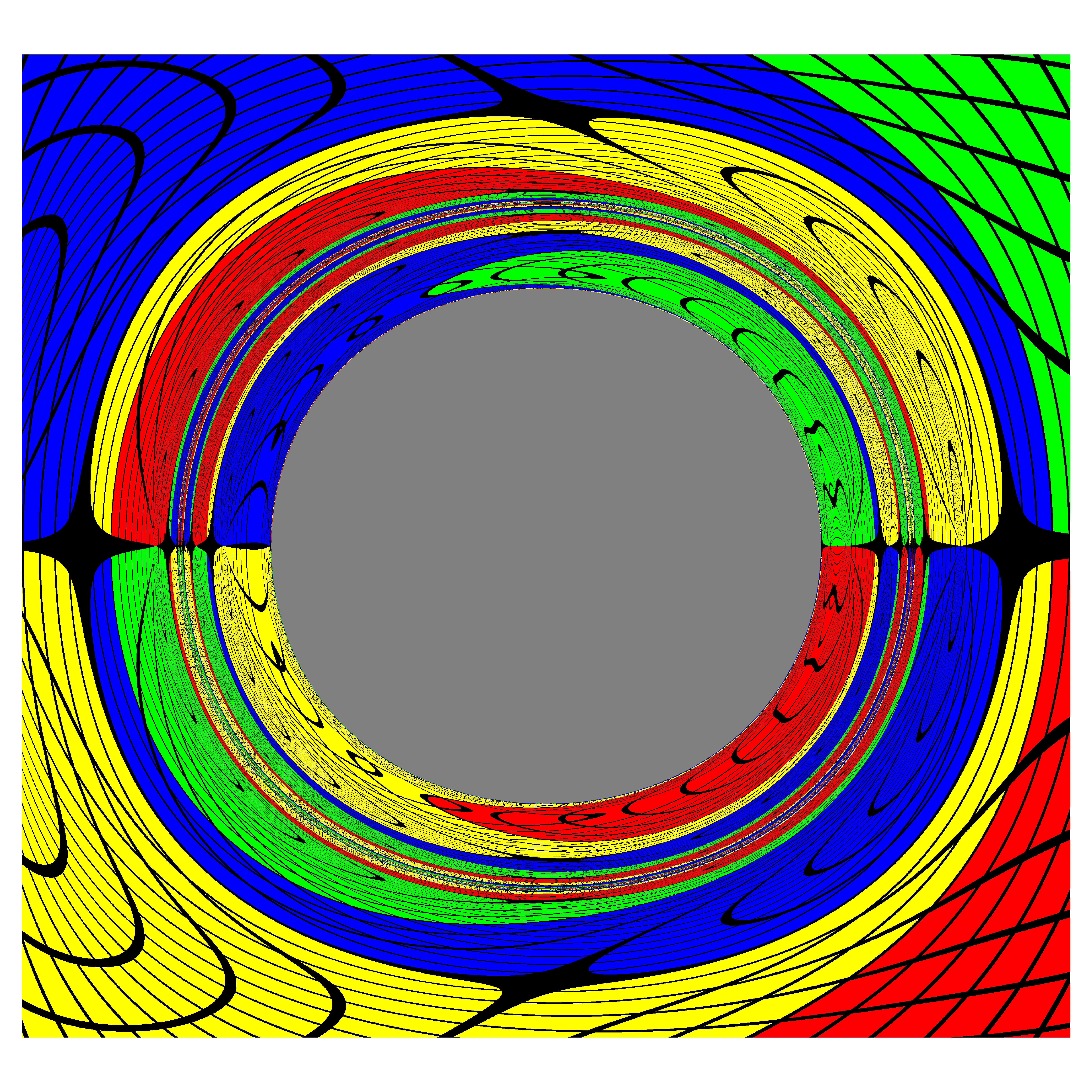}}
		\caption{Images of the black hole with triple photon spheres, which is illuminated by the celestial sphere, viewed by the observers $O$ (left) and $P$ (right). Here, the middle peak is the largest one.}
	\end{figure}
	
	\begin{figure}[htbp]
		\centering
		\subfigure[\ Observer $O$]{
			\includegraphics[scale=0.05]{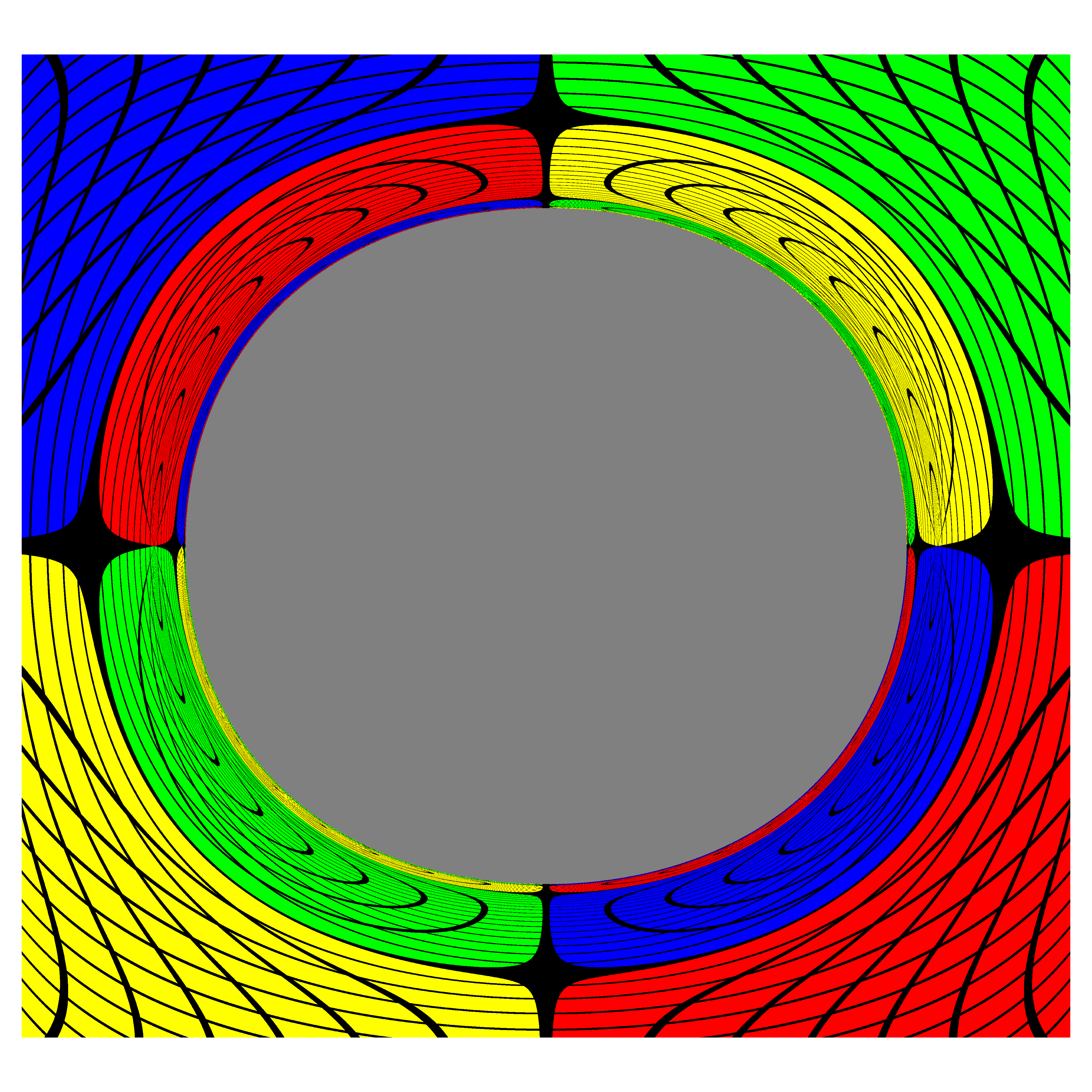}}
		\quad
		\subfigure[\ Observer $P$]{
			\includegraphics[scale=0.067]{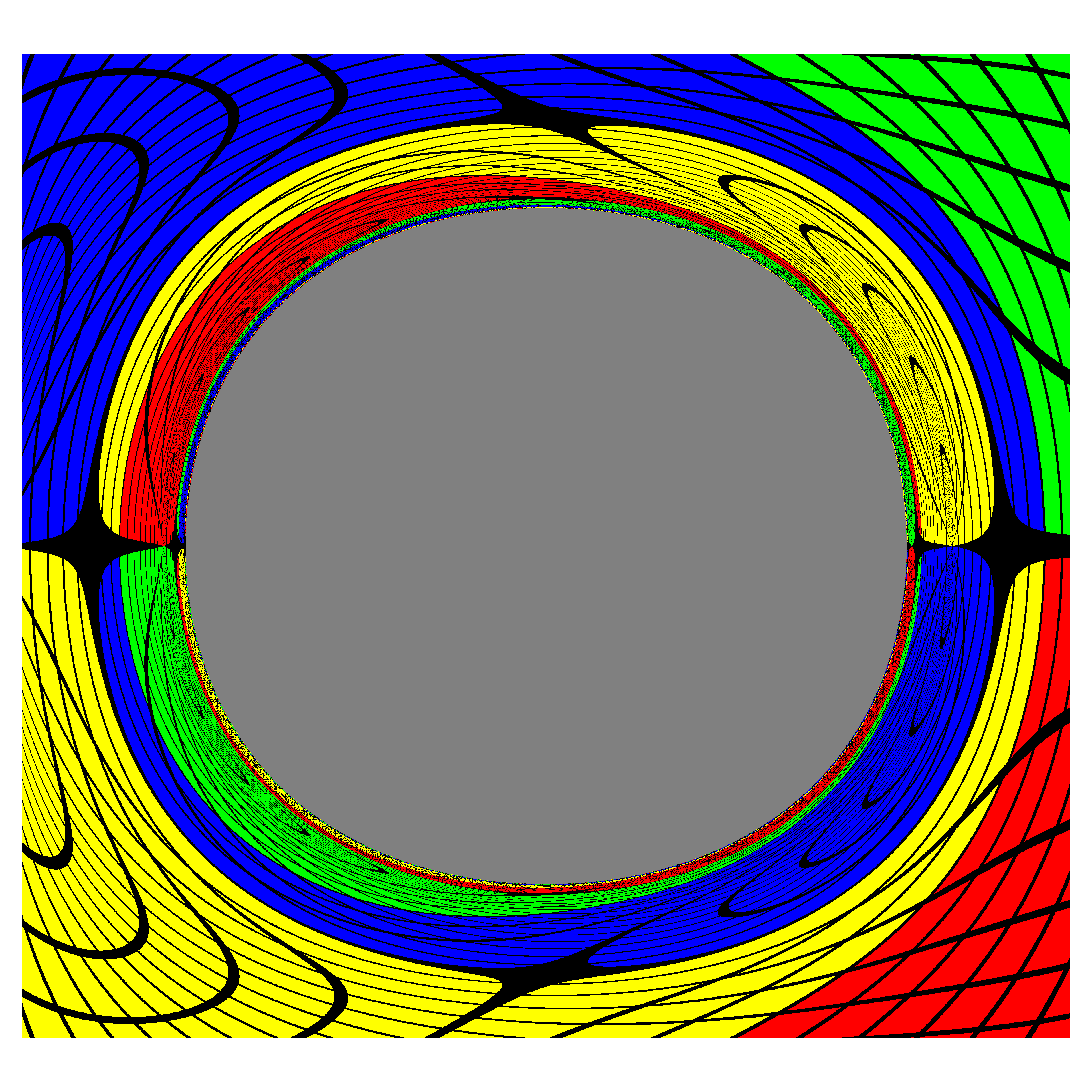}}
		\caption{Images of the black hole with triple photon spheres, which is illuminated by the celestial sphere, viewed by the observers $O$ (left) and $P$ (right). Here, the outer peak is the largest one.}
	\end{figure}
	
	\begin{figure}[htbp]
		\centering
		\subfigure[\ Observer $O$]{
			\includegraphics[scale=0.05]{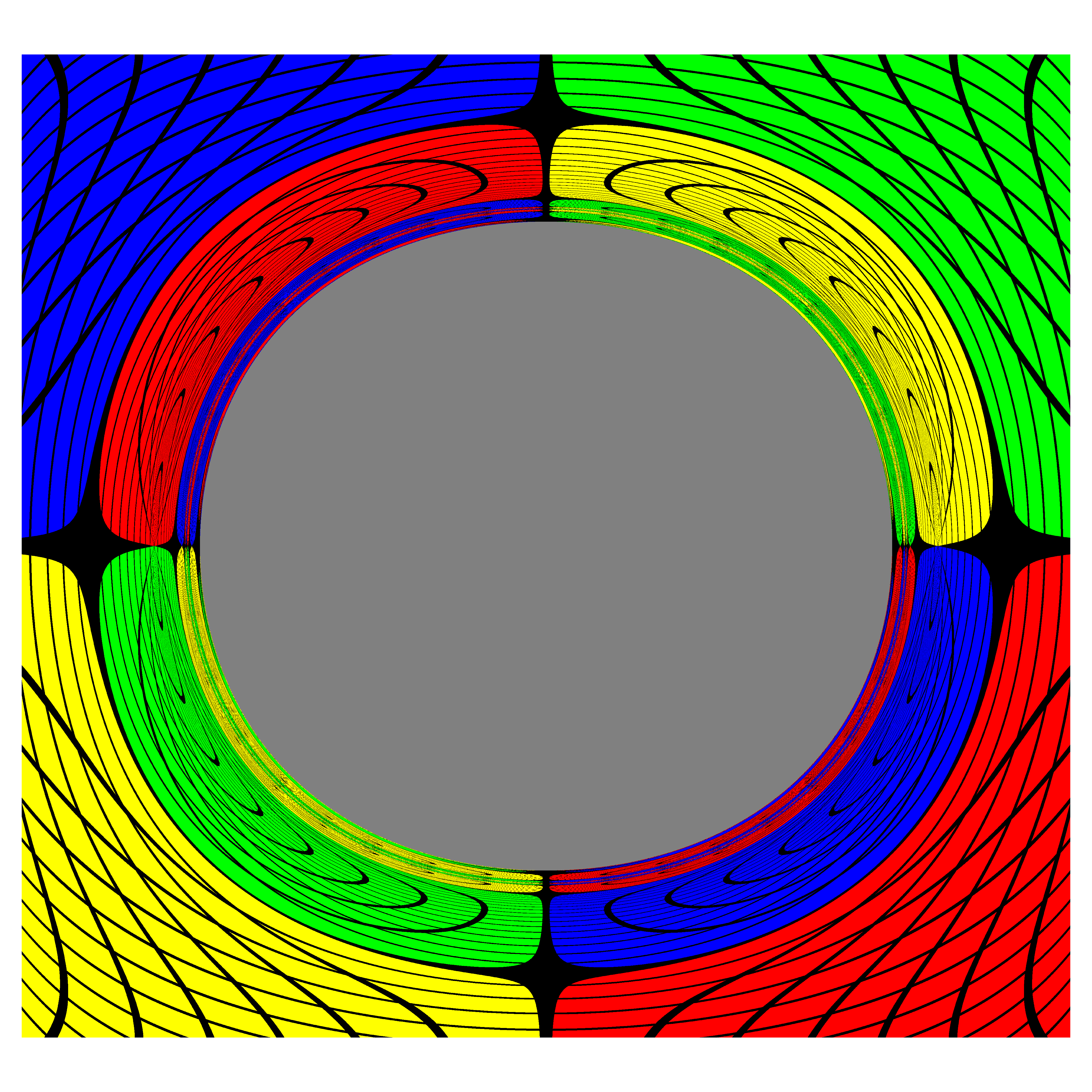}}
		\quad
		\subfigure[\ Observer $P$]{
			\includegraphics[scale=0.05]{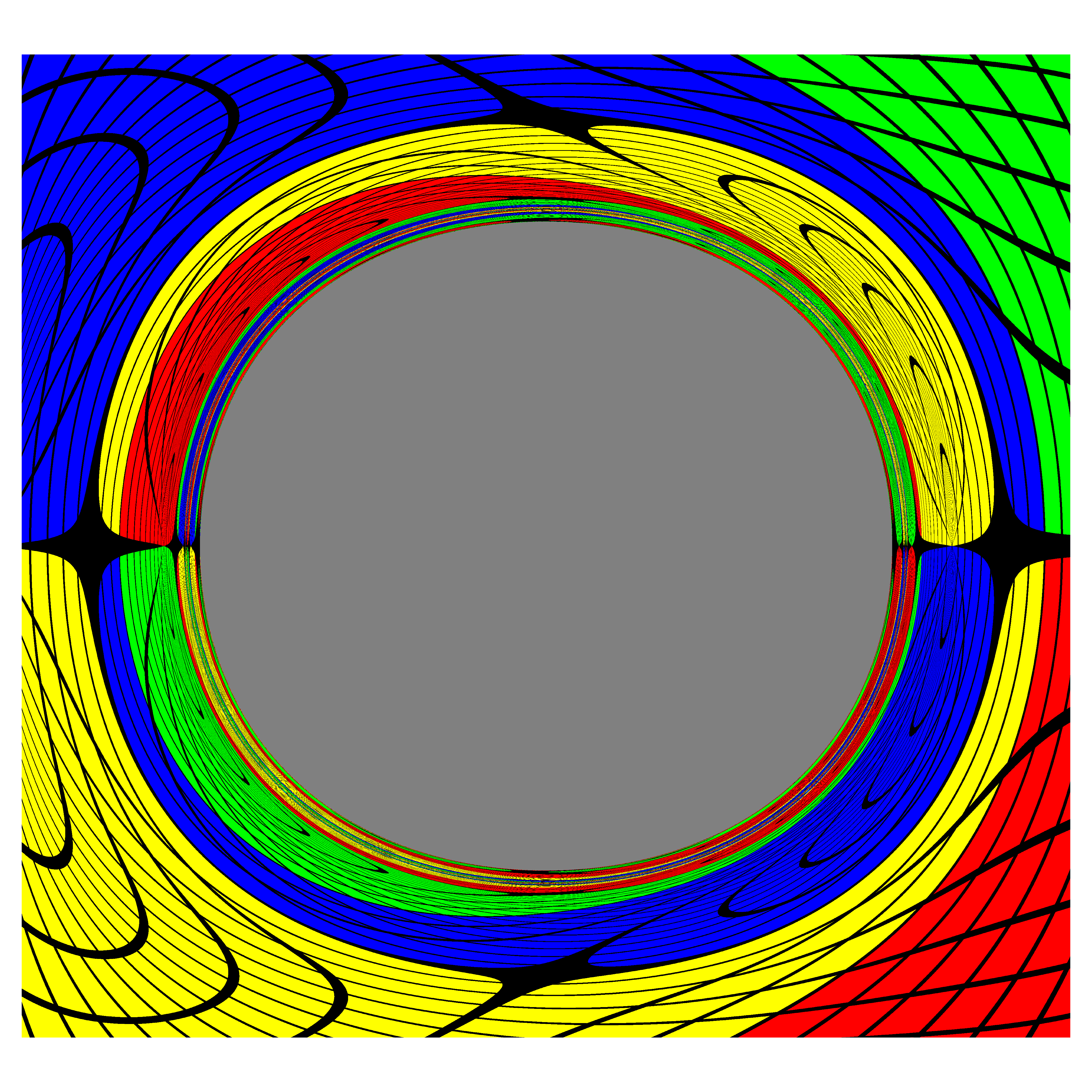}}
		\caption{Images of the black hole with triple photon spheres, which is illuminated by the celestial sphere, viewed by the observers $O$ (left) and $P$ (right). Here, the inner peak is the largest one and the middle peak is the smallest one.}
	\end{figure}

	Whether the effective potential is double-peaked, triple-peaked, or even multi-peaked, we find that it does not always decrease gradually from the inside out. For black hole metrics with specific parameter choices, it is possible for the outer peak to be higher than the inner peak. In such cases, the gravitational influence of the outer photon sphere dominates, effectively concealing the lensing signatures associated with the inner photon spheres. This leads to observational images that resemble those of black holes with fewer photon spheres. The absolute magnitude determines the overall light paths for different impact parameters, while the relative magnitudes of neighboring peaks influence the specific behavior of the photons.
	
	As shown in Figure 8, when the middle peak is the highest, regardless of the size or position of the inner peak, the absolute magnitude of the middle peak ensures that light rays entering the region inside the middle photon sphere cannot be received by the observer. Therefore, in this scenario, the black hole exhibits gravitational lensing effects similar to those of a double-peaked effective potential.
	In Figure 9, when the outermost peak is the highest, the absolute magnitude of this outer peak, irrespective of changes in the middle or inner peaks, conceals all gravitational field information from inside itself. This results in the celestial images observed being consistent with those produced by a black hole with a single-peaked effective potential.
	However, in Figure 10, even though the inner peak is the absolute maximum, the outer peak is still larger relative to the middle peak. Consequently, when considering the outer and middle peaks alone, the outer peak can also obscure the gravitational field information of the middle peak. Together with the inner maximum peak, this configuration creates celestial images similar to those seen around double-peaked black holes.

	\subsection{General rules}
	
	Through analyzing and comparing the ray-tracing images for single-peak, double-peak, and triple-peak effective potentials of black holes, it is evident that each additional peak in the effective potential significantly affects both the formation and the number of higher-order images. By examining the behavior of higher-order ray-tracing images near each critical curve, we observe that when a light ray is influenced by a particular peak in the effective potential, it generally follows one of two distinct trajectories. The first type of trajectory involves the light ray approaching the exterior of a critical curve, looping around it, and then crossing into the interior of this curve before moving towards the location of the next critical curve. The second type involves the light ray first entering the interior region of the critical curve, looping around near the inner edge, and then traveling towards the next critical curve. This explains how black holes with multiple effective potential peaks can simultaneously produce numerous higher-order celestial images near multiple critical curves. Each additional peak introduces two extra trajectory choices for the light rays. In certain cases, the light rays can take these options simultaneously and repeatedly. This leads to the possibility of rays looping multiple times around both the exterior and interior regions of a given critical curve before moving to the next critical curve, thereby generating even higher-order images. These higher-order trajectories are highly sensitive, and even minute variations in the aiming distance can result in drastic changes in the order of the images.
	
	For black hole models with multiple event horizons in Einstein-nonlinear Electrodynamic Theories, the black holes can exhibit a greater number of effective potential peaks, far exceeding just three. However, we can derive a more general rule from our discussions on single-peaked, double-peaked, and triple-peaked effective potentials. Let's assume a black hole with $N$ effective potential peaks, where the outermost peak is labeled as the 1st peak and the innermost as the $N$th peak. Since light rays enter the gravitational field of the black hole from afar with a certain impact parameter, we need to start by comparing from the 1st peak, i.e., the outermost peak, and proceed inward sequentially.
	For instance, when comparing up to the $k$th peak, we need to examine which of the effective potential peaks before the $k$th peak is greater in magnitude relative to it. This helps determine whether the information from the $k$th peak will be hidden by the preceding peaks. If the preceding peaks are larger, the $k$th peak's influence will be concealed. 
	Next, let's assume that the $N$ effective potential peaks ultimately result in $M$ photon spheres. We can then infer the pattern of higher-order images. Based on the conclusions from the triple-peaked black hole case, we can expect $M$ bright photon rings where higher-order images accumulate. These higher-order images will cluster around the critical curves of these $M$ photon spheres. Within each cluster, the higher-order images will be densely packed, while there will be significant spacing between these clusters, making it easier for observers to distinguish them. 
	This general rule implies that the more peaks an effective potential has, the more complex the arrangement of photon spheres and the corresponding distribution of higher-order images will be, leading to intricate and distinguishable gravitational lensing patterns.

	\section{Discussion and conclusion}\label{444}
	
	In this paper, we investigate multi-event horizon black hole models within the framework of Einstein-nonlinear Electrodynamic Theories. Under appropriate parameters, this type of black hole can possess multiple photon spheres \cite{Gao:2021kvr}. Accordingly, we study the gravitational lensing images of point sources and luminous celestial spheres and derive the effects of the number of photon spheres on the resulting lensing images \cite{Guerrero:2022qkh, Guo:2022muy}. For black holes with a single photon sphere, strong gravitational lensing near the photon sphere produces higher-order images of point sources and a series of higher-order images of the celestial sphere outside the shadow boundary. In the case of black holes with two photon spheres, two distinct sets of higher-order images of the celestial sphere accumulate near the two critical curves corresponding to each photon sphere. Additionally, due to differences in the relative sizes of the two effective potentials, dual photon sphere black holes also exhibit celestial images similar to those observed in single photon sphere black holes. For black holes with three photon spheres, the higher-order images near the critical curves become more complex, and due to the more intricate combinations of the relative sizes of the three effective potentials, triple photon sphere black holes also display celestial images similar to those of single and dual photon sphere black holes.
	
	Furthermore, by analyzing the combinations of relative sizes of the effective potentials and the corresponding variations in celestial images for triple photon sphere black holes, we deduce the patterns of celestial image changes for black holes with more than three photon spheres. Although multi-event horizon black hole models in Einstein-nonlinear Electrodynamic Theories can indeed have more than three photon spheres, the derived patterns remain consistent. These unique features provide powerful tools for identifying black holes with multiple photon spheres.
	
	While the size of the black hole shadow may depend on the surrounding astrophysical environment and real black holes may not perfectly match the models described, the trends in gravitational lensing images for multi-photon sphere black holes are likely universal \cite{Guerrero:2022qkh, Guo:2022muy}. Extending our analysis to models that are more aligned with astrophysical realities would be an exciting avenue for exploring the impact of multiple photon spheres on future photon ring observations.

	\section*{Acknowledgement}
	We are grateful to Haitang Yang, Jun Tao, Guangzhou Guo and Yiqian Chen for useful discussions. This work is supported in part by NSFC (Grant No. 11747171), Xinglin Scholars Project of Chengdu University of Traditional Chinese Medicine (Grant no.QNXZ2018050). The authors contributed equally to this work.

\end{document}